# A non-linear fluid suspension model for blood flow

Wei-Tao Wu[1], Nadine Aubry[2], James F. Antaki[3], Mehrdad Massoudi[4*]

1. School of Mechanical Engineering, Nanjing University of Science and Technology, Nanjing, J.S., 210094, China

2. Department of Mechanical and Industrial Engineering, Northeastern University, Boston, MA 02115, USA

3. Nancy E. and Peter C. Meinig School of Biomedical Engineering, Cornell University, Ithaca, NY 14853, USA

4. U. S. Department of Energy, National Energy Technology Laboratory (NETL), Pittsburgh, PA, 15236, USA

Address for correspondence:

Mehrdad Massoudi, PhD,

U. S. Department of Energy,

National Energy Technology Laboratory (NETL),

626 Cochrans Mill Road, P.O. Box 10940,

Pittsburgh, PA. 15236.

Email: Mehrdad.Massoudi@NETL.DOE.GOV

**Abstract**

Motivated by the complex rheological behaviors observed in small/micro scale blood vessels, such as the Fahraeus effect, plasma-skimming, shear-thinning, etc., we develop a non-linear suspension model for blood. The viscosity is assumed to depend on the volume fraction (hematocrit) and the shear rate. The migration of the red blood cells (RBCs) is studied using a concentration flux equation. A parametric study with two representative problems, namely simple shear flow and a pressure driven flow demonstrate the ability of this reduced-order model to reproduce several key features of the two-fluid model (mixture theory approach), with much lower computational cost.

## 1. Introduction

Blood is a dense suspension composed of red blood cells (RBCs), white blood cells (WBCs), platelets in plasma. The hematocrit (Hct) or the RBC volume fraction is about 40-50% for male and 35-45% for female [1]. Blood is responsible for transporting oxygen, removing metabolites and defending the body against infection, etc. The properties and the behavior of blood are intimately related to the safety and efficiency of most blood-wetted medical devices and are also related to many diseases, such as atherosclerosis, cardiovascular and cerebrovascular diseases. This has been the driving force for us to develop mathematical and computational models for improving our capability to quantitatively analyse and predict the response of blood *in vivo* and *in vitro*. Comparing with the costly experimental trial-and-error method, numerical simulations have been used for their high efficiency and low costs in the design of medical devices, especially at the initial phase of a project [2,3].

In large vessels, blood is usually treated as a Navier-Stokes (Newtonian) fluid [4]; while in vessels with the diameter ranging from 20 to 500 microns or when the shear rate is below 100 $s^{-1}$, blood demonstrates some non-Newtonian features, such as stress relaxation, shear-thinning, and plasma skimming [5]. Due to the high volume fraction (hematocrit, about 40%) of the RBCs, the mechanical properties of the whole blood are greatly influenced by the behavior of the RBCs. At low shear rates, due to fibrinogen and large globulins, the RBCs tend to aggregate and form rod-shaped stacks called rouleaux [1,6,7], which increases the blood viscosity; while as the shear rate increases, the aggregation of the RBCs reverses, which leads to a decrease in the viscosity[1]. As the shear rate increases further, the RBCs deform and become elongated, aligned with the flow, further contributing to the shear thinning behavior of blood [8]. In certain conditions the viscosity can also be influenced by the diameter of the blood vessel. [9–11]. Similar to many solid-fluid suspensions, the variation of the hematocrit greatly changes the blood viscosity. It has been observed that as the hematocrit increases, the viscosity increases dramatically [12,13]; and as the hematocrit decreases the shear-thinning property of blood becomes weaker and eventually disappears [14].

Several decades of investigations on blood flow have revealed that in micro-scale vessels/channels, the blood shows outstanding multi-component features. For example, for the capillary vessels with diameters ranging from 50 to 1500 micros, especially when blood flows into a narrow and long tube from a larger tube the hematocrit in the narrow tube is reduced. This phenomenon is known as the Fahraeus effect[15,16]. Kang and Eringen (1976) [17] mention that a similar phenomenon was observed in slurry flows as early as 1958 and is known as the sigma-phenomenon. Another unique feature of blood flow is the rarefaction of the red cells near the walls of smaller branch vessels (known as the Fahraeus–Lindqvist effect, or plasma-skimming effect. Similarly, Segre and Silberberg (1962 a, b) [18,19] observed that starting with a uniform initial distribution of neutrally buoyant particles in a viscous fluid, the particles aggregate at about 0.6 radius from the pipe center [known as the Segre-Silberberg effect].

Motivated by observations of plasma skimming and near-wall platelet enrichment, numerous meso-scale simulations based on different computational methods have been developed. These include Lattice Boltzmann Method - Immersed Boundary Method (LBM-IBM) [20,21], Dissipative Particle Dynamics (DPD) [22,23], Moving Particle Semi-implicit method (MPS) [24], and others. Although the multi-phase aspect of blood has been replicated by numerous meso-scale simulations, their high computational cost limit their use for many realistic engineering scale problems [25]. As alternative methods, multi-component models based on continuum mechanics, ignoring the detail of single cells, can overcome the limitations of high computational cost but still provide useful information, such as the volume fraction (hematocrit) and the velocity fields of RBCs and plasma [2] and the non-uniform distribution of platelet [26].

Flow of a RBCs-plasma (solid-fluid) suspension can be mathematically modeled using different approaches: (1) mixture theory [27], or the averaging method [28], where the two components are treated as two fluid (continua) and are coupled through interaction forces, such as drag force, shear lift force, etc. [2,29]; (2) single phase non-homogenous models where the movement of the RBCs is represented by various concentration fluxes and the transport properties of the suspension are functions of local physical fields. Mixture theory was first presented within the framework of continuum mechanics by Truesdell (1957) [30], and in recent decades it has been widely applied to various applications, reacting immiscible mixtures, growth and remodeling of soft tissues, as well as blood flow [2,31]. The single phase non-homogenous model has been shown to be useful in certain problems including the solid-fluid suspensions [32,33].

One of the most important applications that requires an accurate model of the concentration field of RBCs is thrombosis in blood-wetted medical devices. Virtually any device that comes in contact with blood will experience deposition of blood platelets. The rate of deposition is, in part, governed by the RBCs, which when concentrated will "crowd out" the platelet. If, for example, RBCs concentrate near the centerline of a tube, the platelets will be impelled towards the wall of the tube. This phenomenon is well characterized in capillary tubes and has been well known since the time of Fahraeus, as mentioned above [16]. However, in more complicated geometries, the distribution of RBCs is not straightforward and therefore prompts investigators and developers to rely on numerical simulation. But as a practical matter, it is computationally expensive to perform micro-scale simulations in macro-scale problems. In all but the simplest problems, even mixture theory or averaging methods can be prohibitive. Therefore, if a reduced-order, single phase model for the RBC concentration field is found to be reliable, it could greatly facilitate the numerical simulation of platelet deposition, for example using an auxiliary convection-diffusion approach as suggested by Hund et al. (2009) [34].

In this paper, we develop a suspension model for (whole) blood. In Section 2, we introduce the basic governing equations, and in the following section, we discuss the constitutive relations for the stress tensor and the RBCs transport fluxes. In Sections 4 and 5, we describe and discuss the geometry and the kinematics of problems followed by the numerical results.

## 2. Governing Equations

Let $X$ denote the position of a material point in the fluid. The motion can be represented by

$$x = \chi(X, t) \tag{1}$$

The kinematical quantities associated with the motion are

$$v = \frac{\partial x}{\partial t} \tag{2}$$

$$D = \frac{1}{2}\left(\frac{\partial v}{\partial x} + \left(\frac{\partial v}{\partial x}\right)^T\right) \tag{3}$$

where $v$ is the velocity field, $D$ is the symmetric part of velocity gradient, and $\frac{\partial}{\partial t}$ denotes differentiation with respect to time holding $X$ fixed and superscript '$T$' designates the transpose of a tensor. In the absence of any thermo-chemical and electro-magnetic effects, the governing equations for the flow of a non-linear fluid are the conservation equations for mass, linear momentum, and angular momentum. In addition, we also introduce an equation for conservation of the RBCs. Since we are interested in a purely mechanical system, the energy equation and the entropy inequality are not considered.

## 2.1. Conservation of mass

The conservation of mass is:

$$\frac{\partial \rho}{\partial t} + div(\rho \boldsymbol{v}) = 0 \tag{4}$$

where $\rho = (1 - \phi)\rho_{f0} + \phi\rho_{s0} = \alpha\rho_{f0} + \phi\rho_{s0}$ is the density of the whole blood, $\phi$ is the volume fraction (concentration) of the RBCs, $\rho_{f0}$ and $\rho_{s0}$ are the pure density of plasma and RBCs in the reference configuration (before mixing). $\partial/\partial t$ is the derivative with respect to time, and $div$ is the divergence operator. For an isochoric motion we have $div\ \boldsymbol{v} = 0$ [for a recent discussion on isochoric motions see Rajagopal, et al. (2015) [35]].

## 2.2. Conservation of linear momentum

Let $\boldsymbol{T}$ represent the Cauchy stress tensor. Then the balance of the linear momentum is:

$$\rho \frac{d\boldsymbol{v}}{dt} = div\boldsymbol{T} + \rho \boldsymbol{b} \tag{5}$$

where $\frac{d\boldsymbol{v}}{dt} = \frac{\partial \boldsymbol{v}}{\partial t} + (grad\boldsymbol{v})\boldsymbol{v}$ and $\boldsymbol{b}$ stands for the body force. The conservation of angular momentum implies that in absence of couple stresses the Cauchy stress tensor is symmetric, that is $\boldsymbol{T} = \boldsymbol{T}^T$.

## 2.3. Conservation of RBCs (particles) concentration (Convection-diffusion)[1]

The equation for the conservation of the RBCs concentration is [32],

---

[1] In their study of a pulsatile flow of a non-linear chemically- reacting fluid, Bridges and Rajagopal (2006) [61] assumed that the viscosity of the fluid, not only depends on the concentration of the species (or constituent whose behavior is governed by a convection-reaction-diffusion equation), but also on the velocity gradient, in a properly frame-invariant form. They [61] used the following equation:

$$\frac{\partial \phi}{\partial t} + div(\phi \boldsymbol{v}) = f$$

where $\phi$ is the concentration and $f$ is a constitutive parameter. We can see the similarity between the above two approaches. They assumed

$$f = -div\boldsymbol{W}$$

where $\boldsymbol{W}$ is a flux vector, related to the chemical reactions occurring in the fluid and is assumed to be given by a constitutive relation similar to the Fick's assumption, namely

$$\boldsymbol{W} = -K_1 \nabla C$$

where $K_1$ is a material parameter which in general is not constant. Bridges and Rajagopal (2006) [61] assumed that $K_1$ is a scalar-valued function of (the first Rivlin-Ericksen tensor) $\boldsymbol{A}_1$

$$K_1 = K_1(\boldsymbol{A}_1) = k\|\boldsymbol{A}_1{}^2\|$$

where $k$ was assumed constant and $\|\cdot\|$ denotes the trace-norm. This approach, even though used for a single continuum complex fluid, is really based on a constrained mixture theory approach [see Humphrey and Rajagopal (2002) [62]] indicating that, at each point in the mixture, the host fluid coexists with the constituent, flowing in such a way that the two components are constrained to move together. In such an approach, one does not need to study the complicated system of equations for the multi-component mixtures.

$$\frac{\partial \phi}{\partial t} + \boldsymbol{v}.\,grad\boldsymbol{\phi} = -div\boldsymbol{N} \qquad (6)$$

where the first term on the left hand side denotes the rate of accumulation of RBCs and the second term denotes the convected RBCs flux (where $\frac{\partial \phi}{\partial x_i}$ or $grad\boldsymbol{\phi}$ denotes the gradient of the concentration). The term on the right side denotes the diffusive RBCs flux $\boldsymbol{N}$ which is composed of fluxes related to the Brownian motion, and the variation of the interaction frequency and the viscosity [32,36]. For a detailed discussion of the convection-diffusion equation see Rajagopal, et al., (2010) [37]. In the next section, we discuss the constitutive equations used in our study.

## 3. Constitutive Equations

Most complex fluids are mixtures composed of different components (phases). At times, these fluids can be assumed to behave as a single continuum suspension with non-linear material properties. [See, for example, Macosko (1994) [38]]. In this case, one studies the global or macroscopic properties of the suspension, for example, the velocity or temperature fields of the whole suspension. In other applications, if one needs to know the details of the velocity, concentration, temperature fields for *each* constituent (phase), then a multi-component (phase) approach should be used [See Rajagopal and Tao (1995) [39], Massoudi (2008, 2010) [31,40]]. Examples of complex fluids whereby both approaches can be used are coal-slurries, biological fluids such as blood and synovial fluid, and chemically- reacting fluids. Using the single component approach, there are primarily two distinct, yet complementary methods that can be applied to constitutive modeling of blood. One approach is based on ideas in the micro-continuum or the structured continua theories [see Eringen (1991, 2005) [41,42], Ariman et. al (1973) [43]] whereby additional balance equations are required. Alternatively, blood can be viewed as a suspension and modeled using the techniques of non-Newtonian fluid mechanics. The latter approach is adopted in this study. A general approach to formulate constitutive relations has recently been provided by Rajagopal and Saccomandi (2016) [44].

### 3.1. Stress tensor

Although blood plasma behaves as a Newtonian fluid, the RBCs (which comprise approximately 45% of the volume of normal human blood) introduces shear-thinning behavior. Therefore, whole blood must be modeled as a non-Newtonian fluid [see Anand and Rajagopal (2004) [45]]. Most of the previous rheological models of blood introduce non-Newtonian behavior through: (i) shear-thinning (e.g., using power law or Oldroyd-type models), and/or (ii) yield stress (e.g., using a Casson model or Herschel-Bulkley-type models). Due to the deformability of the RBCs and their shape, it is feasible that blood would exhibits normal stress effects, such as die-swell or rod-climbing, exhibited in polymeric materials [see for example, Macosko (1994) [38]]. However, there is not much experimental evidence to suggest this is a significant property of blood. It is also feasible that blood exhibit viscoelastic properties, due to the rheological mechanisms for reversible storage of energy (e.g. due to the deformation and aggregation/disaggregation of the RBCs). Under certain conditions, there is some experimental evidence to support this assumption. [See Thurston (1972) [46], Chien et al (1975) [47]]. One of the successful models which has been able to capture the shear-thinning behavior of blood over a wide range of shear rates is a generalized three-constant Oldroyd-B model proposed by Yeleswarapu (1996) [48] and Yeleswarapu et al. (1998) [10].

According to many experimental observations, blood shear viscosity depends on the shear rate and the volume fraction [14,46,49]. Therefore, we assume blood can be modeled as,

$$\boldsymbol{T} = -p\boldsymbol{I} + \mu(\phi, \dot{\gamma})\boldsymbol{D} \tag{7}$$

where $\mu$ is the viscosity of the (whole) blood depending on the volume fraction and the shear rate ($\dot{\gamma} = \sqrt{2tr(\boldsymbol{D}^2)}$), $\boldsymbol{I}$ is the identity tensor, $p$ is the pressure. Note that we have ignored the viscoelastic properties or the yield-stress of blood. We also assume that the viscosity is given by a Carreau-type model, where the viscosity is assumed to vary with the shear rate. When the shear rate is close to zero, the viscosity approaches a lower limit, $\mu_0$; when the shear rate is close to infinity, the viscosity approaches an upper limit, $\mu_\infty$. Thus, $\mu$ is given by [as suggested by Yeleswarapu (1994) [50]].

$$\mu = \mu_\infty(\phi) + \left(\mu_0(\phi) - \mu_\infty(\phi)\right)\frac{1 + \ln(1 + k\dot{\gamma})}{1 + k\dot{\gamma}} \tag{8}$$

where $\mu_0(\phi)$ and $\mu_\infty(\phi)$ are the viscosities when the shear rate approaches zero and infinity, respectively, and $k$ is the shear-thinning parameter. The values or correlations of $\mu_0(\phi)$, $\mu_\infty(\phi)$ and $k$ are determined by experiments measurement. We further assume a volume-fraction dependency of $\mu$, such that,

$$\mu = \left[\mu_\infty + (\mu_0 - \mu_\infty)\frac{1 + \ln(1 + k\dot{\gamma})}{1 + k\dot{\gamma}}\right](c_{\mu0} + c_{\mu1}\phi + c_{\mu2}\phi^2) \tag{9}$$

According to Massoudi (2008) [31] and Batchelor and Green (1972) [51], we assume $c_{\mu0} = 1$, $c_{\mu1} = 2.5$ and $c_{\mu2} = 7.6$.

## 3.2. Fluxes due to the motion of the RBCs

The motion of RBCs can influence the distribution of platelets due to the non-homogeneity of their concentration field. Regions in which RBCs are most concentrated may "crowd out" the platelets. To model the fluxes due to the motion of RBCs, we will use the concept of a convection-diffusion equation.

For the particle flux, we use the approach of Phillips et al, [29], who suggested that the particles flux $\boldsymbol{N}$ , in general, can be due to the Brownian motion, turbulent diffusivity, particles interactions, gravity, etc. [32,33,36]. We ignore the Brownian motion due to the (large) size of RBCs and we also assume that the flow is laminar; thus $\boldsymbol{N}$ is given by:

$$\boldsymbol{N} = \boldsymbol{N}_c + \boldsymbol{N}_\mu + \boldsymbol{N}_g \tag{10}$$

where $\boldsymbol{N}_c$, $\boldsymbol{N}_\mu$ and $\boldsymbol{N}_g$ are the contributions to the flux due to particles collision, spatial variations in the viscosity and gravity, respectively. Based on the ideas developed in Phillips et al, [32], we assume $\boldsymbol{N}_c$ and $\boldsymbol{N}_\mu$ are given by:

$$\boldsymbol{N}_c = -a^2 \phi K_c \nabla(\dot{\gamma}\phi) \tag{11}$$

$$\boldsymbol{N}_\mu = -a^2 \phi^2 \dot{\gamma} K_\mu \nabla(\ln \mu) \tag{12}$$

$$\dot{\gamma} = (2\boldsymbol{D}_{ij}\boldsymbol{D}_{ij})^{1/2} = (\Pi)^2 \tag{13}$$

where $a$ is the characteristic reference length (e.g. radius), and $K_c$ and $K_\mu$ are empirically-determined coefficients. Based on the work of Acrivos et al. [52], we model $\boldsymbol{N}_g$ as,

$$\boldsymbol{N}_g = \frac{2}{9} \phi \frac{(1-\phi)}{\mu(\phi,\dot{\gamma})} a^2 (\rho_{s0} - \rho_{f0})\boldsymbol{g} \tag{14}$$

The above equation has been used in several applications, such as the falling flow of a thin film [53]. Other potential models for $\boldsymbol{N}_g$, are given in [54,55].

## 4. Geometry and the kinematics of the flows

In this paper, we study two simple flows: (1) a simple shear flow and (2) a pressure driven flow. The geometry and the kinematics of the two boundary value problems are shown in Figure 1. These are the basis for a parametric study of the dimensionless numbers in the model, above [56–59].

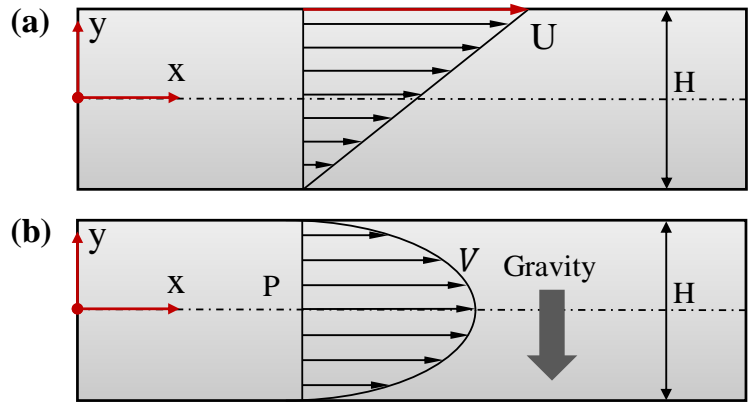

Figure 1 Schematic of the two benchmark problems. (a) Simple shear flow between two infinite plates, with the upper plate moving. (b) Pressure driven flow.

The vectorial (expanded) form of the conservation of linear momentum is:

$$\rho \left[ \frac{\partial \boldsymbol{v}}{\partial \tau} + (grad\boldsymbol{v})\boldsymbol{v} \right]$$
$$= -grad(p) + div\left( \left( \mu_\infty(\phi) + (\mu_0(\phi) - \mu_\infty(\phi)) \frac{1 + ln(1 + k(2tr\boldsymbol{D}^2)^{1/2})}{1 + k(2tr\boldsymbol{D}^2)^{1/2}} \right) \boldsymbol{D} \right) + \rho\boldsymbol{b} \tag{15}$$

The RBCs concentration equation is:

$$\frac{\partial \phi}{\partial t} + \boldsymbol{v}.grad\phi$$
$$= div\left( a^2 \phi K_c \nabla(\dot{\gamma}\phi) + a^2 \phi^2 \dot{\gamma} K_\mu \nabla(\ln \mu(\phi,\dot{\gamma})) - \frac{2}{9}\phi \frac{(1-\phi)}{\mu(\phi,\hat{\gamma})} a^2 \left(\rho_{s0} - \rho_{f0}\right) \boldsymbol{g} \right) \tag{16}$$

The above governing equations can be non-dimensionalized as:

$$\rho \left[ \frac{\partial \boldsymbol{V}}{\partial \tau} + (grad\boldsymbol{V})\boldsymbol{V} \right]$$
$$= -gradP + [B_{31}(1 + 2.5\phi + 7.6\phi^2) + B_{32}(1 + 2.5\phi + 7.6\phi^2)\Pi]div\boldsymbol{D}$$
$$+ B_{31}\boldsymbol{D}grad(1 + 2.5\phi + 7.6\phi^2)$$
$$+ B_{32}\boldsymbol{D}[\Pi grad(1 + 2.5\phi + 7.6\phi^2) + (1 + 2.5\phi + 7.6\phi^2)grad\Pi]$$
$$+ \left(G_f(1-\phi) + G_s\phi\right)\boldsymbol{g} \tag{17}$$

$$\frac{\partial \phi}{\partial \tau} + \boldsymbol{V}\frac{\partial \phi}{\partial \boldsymbol{X}} = div\left( J_c\phi\nabla(\Gamma\phi) + J_\mu\Gamma\phi^2\frac{\nabla\mu}{\mu} - J_g\frac{\phi(1-\phi)}{(B_{31} + B_{32}\Pi)(1 + 2.5\phi + 7.6\phi^2)}\boldsymbol{g} \right) \tag{18}$$

in which the following non-dimensional parameters are defined by:

$$Y = \frac{y}{H_r}; \ X = \frac{x}{H_r}; \ Z = \frac{z}{H_r}; \ \boldsymbol{V} = \frac{\boldsymbol{v}}{u_0}; \ \tau = \frac{tu_0}{H_r}; \ \Gamma = \frac{H_r\dot{\gamma}}{u_0};$$

$$\boldsymbol{g}^* = \frac{\boldsymbol{g}}{g}; \ \rho^* = \frac{\rho}{\rho_r}; \ \rho_f^* = \frac{\rho_f}{\rho_r}; \ \rho_s^* = \frac{\rho_s}{\rho_r};$$

$$\text{div}^*(\cdot) = H_r\text{div}(\cdot); \ \text{grad}^*(\cdot) = H_r\text{grad}(\cdot); \ \boldsymbol{L}^* = \text{grad}^*\boldsymbol{V}$$

$$\boldsymbol{L}^* = \text{grad}^*\boldsymbol{V}; \ \boldsymbol{D}^* = \frac{1}{2}[\text{grad}^*\boldsymbol{V} + (\text{grad}^*\boldsymbol{V})^T]; \tag{19}$$

$$Fr = \frac{u_0^2}{H_r g}; \ P = \frac{p}{\rho_0 u_0^2}; B_{31} = \frac{\mu_\infty}{\rho_0 u_0 H_r}; \ B_{32} = \frac{\mu_0 - \mu_\infty}{\rho_0 u_0 H_r}; \ \Pi = \frac{1 + ln(1 + \bar{k}\ \Gamma)}{1 + \bar{k}\ \Gamma};$$

$$\bar{k} = \frac{ku_0}{H_r}; \ G_f = \frac{\rho_f^*}{Fr}; G_s = \frac{\rho_s^*}{Fr}; \ J_c = \frac{a^2K_c}{H_r^2}; \ J_\mu = \frac{a^2K_\mu}{H_r^2}; J_g = \frac{2a^2(\rho_{s0} - \rho_{f0})g}{9u_0^2\rho_0 H_r}.$$

where $H_r$ is a reference length, for example, the distance between the two plates, and $u_0$ is a reference velocity. (The asterisks in equation (15) have been omitted for simplicity.) Among the above dimensionless numbers, $Fr$ is the Froude number, $\bar{k}$ is a parameter related to the shear-thinning effects, $B_{31}$ and $B_{32}$ are related to the viscous effects (similar to the Reynolds number), $G_f$ and $G_s$ are related to gravity, and $J_c$, $J_\mu$ and $J_g$ are related to the RBCs flux.

We assume that the flow is steady and fully developed:

$$\boldsymbol{V} = U(Y)\boldsymbol{e}_x; \ \phi = \phi(Y) \tag{20}$$

where $\boldsymbol{e}_x$ is the unit vector in the $x$ direction. Using Equation (20), we notice that,

$$\boldsymbol{D} = \frac{1}{2}(grad\boldsymbol{v} + (grad\boldsymbol{v})^T) = \frac{1}{2}\begin{bmatrix} 0 & U' & 0 \\ U' & 0 & 0 \\ 0 & 0 & 0 \end{bmatrix} \tag{21}$$

With equation (20), the equation for the conservation of mass is automatically satisfied. In other words, the motion is isochoric,

$$tr\boldsymbol{D} = div\boldsymbol{V} = 0 \tag{22}$$

Furthermore, $\boldsymbol{V}.grad(.)$ is zero, therefore, the concentration equation (16) simplifies to,

$$0 = div\left(J_c\phi\nabla(\Gamma\phi) + J_\mu\Gamma\phi^2\frac{\nabla\mu}{\mu} - J_g\frac{\phi(1-\phi)}{(B_{31} + B_{32}\Pi)(1 + 2.5\phi + 7.6\phi^2)}\boldsymbol{g}\right) \tag{23}$$

Also, to ensure that there is no RBC penetration through the solid boundary, the RBC flux should be zero [32]:

$$0 = \boldsymbol{n} \cdot \left(J_c\phi\nabla(\Gamma\phi) + J_\mu\Gamma\phi^2\frac{\nabla\mu}{\mu} - J_g\frac{\phi(1-\phi)}{(B_{31} + B_{32}\Pi)(1 + 2.5\phi + 7.6\phi^2)}\boldsymbol{g}\right)\Bigg|_{wall} \tag{24}$$

Integrating equation (23) subject to the boundary condition (24), we have,

$$0 = J_c\phi\nabla(\Gamma\phi) + J_\mu\Gamma\phi^2\frac{\nabla\mu}{\mu} - J_g\frac{\phi(1-\phi)}{(B_{31} + B_{32}\Pi)(1 + 2.5\phi + 7.6\phi^2)}\boldsymbol{g} \tag{25}$$

The above equation implies that the total flux should be zero everywhere in the flow.

Substituting equations (20) and (22) into equations (15) and (25), we obtain three non-linear ordinary differential equations. The momentum equation in the x-direction is:

$$\Delta P + (1 + 2.5\phi + 7.6\phi^2)\left[B_{31} + B_{32}\Pi + \left(B_{32}\frac{1}{\Gamma}\frac{d\Pi}{d\Gamma}U'^2\right)\right]U'' \\ + (2.5 + 15.2\phi)\phi'(B_{31} + B_{32}\Pi)U' = 0 \tag{26}$$

where $\Delta P = -\frac{\partial P}{\partial X}$. Notice that $\Delta P = 0$, for simple shear flow problem as shown in Figure 1 (a). The momentum equation in the y-direction is:

$$\frac{\partial P}{\partial Y} + G_f(1-\phi) + G_s\phi = 0 \tag{27}$$

The RBCs flux (concentration) equation is:

$$J_c \phi (\Gamma \phi)'$$

$$+ J_\mu \Gamma \phi^2 \frac{\left( \left( B_{32} \frac{1}{r} \frac{d\Pi}{dr} U' U'' \right) \left( 1 + 2.5\phi + 7.6\phi^2 \right) + [B_{31} + B_{32}\Pi](2.5 + 15.2\phi)\phi' \right)}{[B_{31} + B_{32}\Pi]\left( 1 + 2.5\phi + 7.6\phi^2 \right)}$$

$$+ J_g \frac{\phi(1 - \phi)}{(B_{31} + B_{32}\Pi)\left( 1 + 2.5\phi + 7.6\phi^2 \right)} = 0 \tag{28}$$

where $U$ is the velocity and $\boldsymbol{g}^* = (0, -1, 0)$. Equation (27) implies a pressure distribution along the y-direction is balanced by the RBCs distribution. From equations (26) and (28), it can be seen that we need two boundary conditions for $U$ and one boundary condition for $\phi$.

For the case of simple shear flow [Figure 1 (a)], we assume the no-slip condition for the lower plate and apply a constant velocity for the moving (upper) plate:

$$U(Y = 1) = 1; U(Y = -1) = 0 \tag{29}$$

For the pressure driven flow problem [Figure 1 (b)], we use the no-slip condition at both boundaries for the velocity:

$$U(Y = 1) = U(Y = -1) = 0 \tag{30}$$

For the volume fraction, $\phi$, an appropriate boundary condition may be given as an average volume fraction (hematocrit) defined in an integral form:

$$\int_{-1}^{1} \phi \, dY = N \tag{31}$$

Other possible boundary conditions for $\phi$ could be given at $Y = -1$:

$$\phi \to \Theta \text{ as } Y \to -1 \tag{32}$$

where $\Theta$ is the value of the volume fraction at the boundary. The condition of the average volume fraction is used in this paper.

## 5. Results and Discussion

The system of the non-linear ordinary differential equations (26) and (28) subject to the boundary conditions given by equations (29) - (31) are solved using the MATLAB solver bvp4c [60]. The step size is automatically adjusted by the solver. The default relative tolerance for the maximum residue is 0.001. The boundary conditions for the average volume fraction (hematocrit) are numerically satisfied by using the shooting method.

### 5.1. Simple shear flow

We first perform a parametric study for the case of simple shear flow, an idealization of the kinematics of a shear viscometer. From Figure 2 (a), it can be seen that the effect of $\bar{k}$ (the shear-thinning effect of viscosity) on the velocity distribution is negligible, and all the velocity curves are almost linear; however, the non-linearity of the RBCs volume fraction becomes more pronounced as $\bar{k}$ increases: more RBCs tend to concentrate near the bottom plate. This can be explained by the increased mobility of RBCs to migrate downwards under the influence of gravity due to the decreased viscosity. Figure 2 (b) shows that as $B_{31}$ increases, the velocity profile becomes more linear. This can be understood by considering the definition of the dimensionless numbers $B_{31}$ and $B_{32}$ in equation (19), in which increasing of $B_{31}$ causes the suspension to become more shear independent. Figure 2 (c) indicates that the effect of $B_{32}$ on the velocity is not significant; however, the effect on RBCs volume fraction distribution is to becomes more uniform as $B_{32}$ increases, that is, as viscosity of the blood increases. This is consistent with the effect of $\bar{k}$. As indicated by equation (14), the RBCs flux due to gravity is negatively proportional to the blood viscosity.

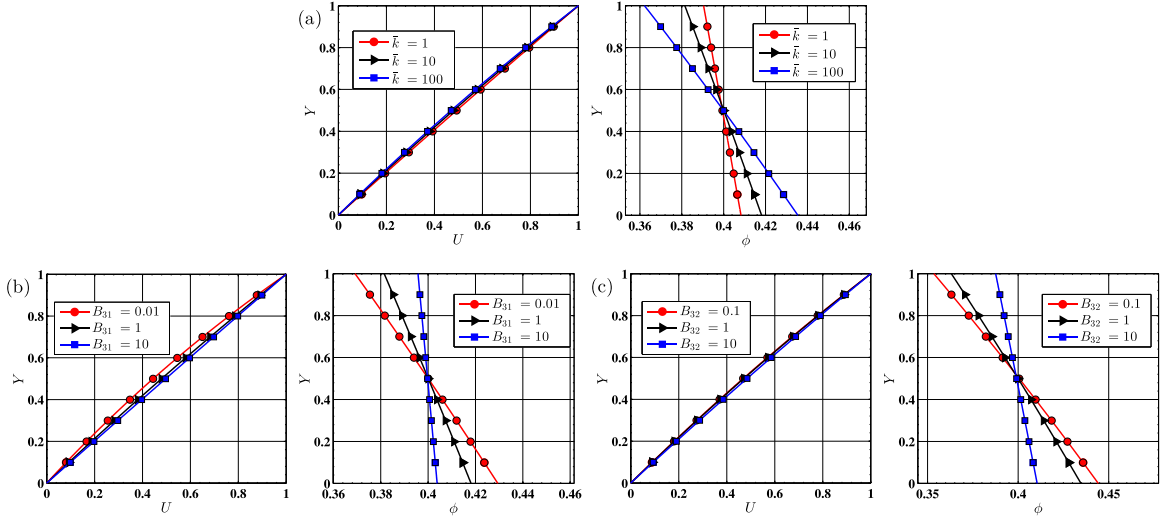

Figure 2. Parametric studies for the shear-thinning viscosity case. (a) Effect of $\bar{k}$ on the velocity profile (left) and the volume fraction profile (right), with $B_{31} = 1$, $B_{32} = 5$. (b) Effect of $B_{31}$ on the velocity profile (left) and the volume fraction profile (right), with $\bar{k} = 10$, $B_{32} = 5$. (c) Effect of $B_{32}$ on the velocity profile (left) and the volume fraction profile (right), with $\bar{k} = 10$, $B_{31} = 1$. We assumed, $J_g = 0.5$, $J_c = 1$, $J_\mu = 1$, $\bar{\phi} = 0.4$.

In Figure 3 we present the impact of the RBCs particles fluxes, $J_c$, $J_\mu$ and $J_g$. As shown in Figure 3 (a), as $J_g$ increases, that is, as the effects of gravity becomes stronger, the RBCs distributions becomes more non-uniform; therefore, the velocity profiles become more non-linear, which can be attributed to the variation of the viscosity along the Y-direction. It is worth pointing out that when $J_g = 0$, the volume fraction profiles are uniform, indicates that in the current problem the non-uniform distribution of the RBCs is mainly due to the gravity. When $J_g = 0$, the volume fraction, shear rate, and viscosity distributions along the Y-direction are all uniform, and there are no fluxes according to equations (11) and (12). Figure 3 (b) and Figure 3 (c) indicate that as $J_c$ increases or $J_\mu$ decreases, the RBCs distribution becomes more non-uniform. This is consistent with our previous studies (for dense suspensions and drilling fluids) [33,36], where it was found that $N_c$ ($J_c$) is responsible for the non-uniform distribution of particles and $N_\mu(J_\mu)$ usually causes the opposite effect as $N_c$ ($J_c$). For velocity profiles, in the range of the values studied here, the effect of $J_c$ is negligible; while as $J_\mu$ decreases, the velocity curves show obvious non-linearity.

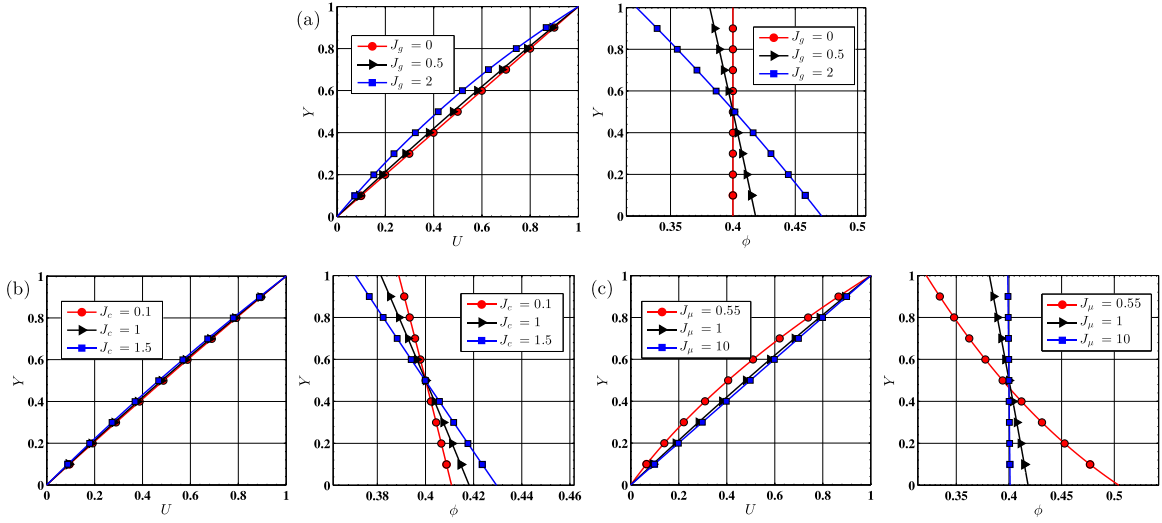

Figure 3. Parametric studies of the RBCs transport fluxes. (a) Effect of $J_g$ on the velocity profile (left) and the volume fraction profile (right), with $J_c = 1, J_\mu = 1$. (b) Effect of $J_c$ on the velocity profile (left) and the volume fraction profile (right), with $J_g = 0.5, J_\mu = 1$. (c) Effect of $J_\mu$ on the velocity profile (left) and the volume fraction profile (right), with $J_g = 1, J_c = 1$. For the above numerical studies, $\bar{k} = 10, B_{31} = 1, B_{32} = 5, \bar{\phi} = 0.4$.

## 5.2. Pressure driven flow

The second problem we consider is the pressure driven flow between two flat horizontal plates. In this problem, we ignore the effect of gravity, and we also apply a symmetry boundary condition at the centerline. Figure 4 shows the shear-thinning effect of the viscosity. Overall the RBCs tend to concentrate near the center, resulting from the greater shear rates near the walls. The velocity profiles all exhibit a nearly parabolic profile. These patterns of the velocity and the volume fraction profiles are consistent with those predicted by our previous two-fluid model [2,29]. Figure 4 (a) shows that increasing $\bar{k}$ (decreasing blood viscosity due to shear thinning) leads to a greater velocity and a less non-uniform distribution of the RBCs. Significant contributions to the non-uniform distribution of the RBCs is the decreasing flux, $\mathbf{N}_\mu$ (related to the viscosity), and the increasing flux, $\mathbf{N}_c$ (related to the shear rate). Figure 4 (b) and Figure 4 (c) indicate that as $B_{31}$ and $B_{32}$ increase, which increases blood viscosity, the velocity decreases, and the volume fraction becomes more uniform. Figure 5 shows the effect of $\mathbf{N}_\mu(J_\mu)$ and $\mathbf{N}_c(J_c)$ on the velocity and the volume fraction distribution. Consistent with the simple shear flow, $\mathbf{N}_c(J_c)$ contributes to the non-uniform distribution of the RBCs while $\mathbf{N}_\mu(J_\mu)$ plays the opposite role. As $J_c$ increases and $J_\mu$ decreases, the velocity increases slightly. The influence of bulk volume fraction (the hematocrit) is displayed in Figure 6. Here, we see that slight variation of $\bar{\phi}$ has a dramatic effect on both the velocity and the volume fraction profiles.

As hematocrit $\bar{\phi}$ increases, the velocity decreases; and the volume fraction distribution is more blunt and uniform. This is a consequence of the dependency of both viscosity and RBC fluxes on volume fraction.

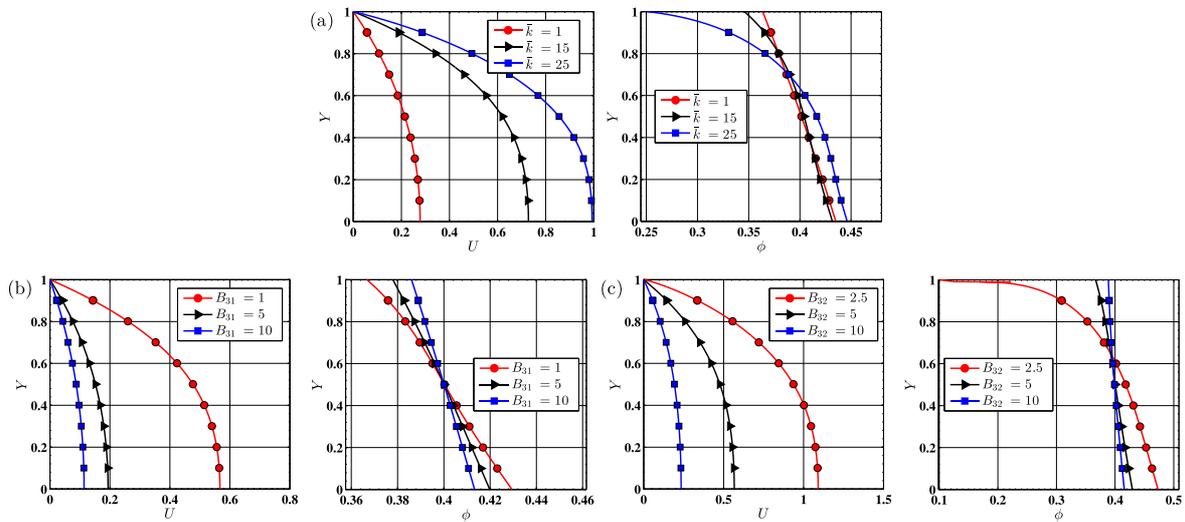

Figure 4. Parametric studies on the shear-thinning viscosity. (a) Effect of $\bar{k}$ on the velocity profile (left) and the volume fraction profile (right), with $B_{31} = 1$, $B_{32} = 5$. (b) Effect of $B_{31}$ on the velocity profile (left) and the volume fraction profile (right), with $\bar{k} = 10$, $B_{32} = 5$. (c) Effect of $B_{32}$ on the velocity profile (left) and the volume fraction profile (right), with $\bar{k} = 10$, $B_{31} = 1$. For the above numerical studies, $J_c = 0.5$, $J_\mu = 1$, $\bar{\phi} = 0.4$, $\Delta P = 10$.

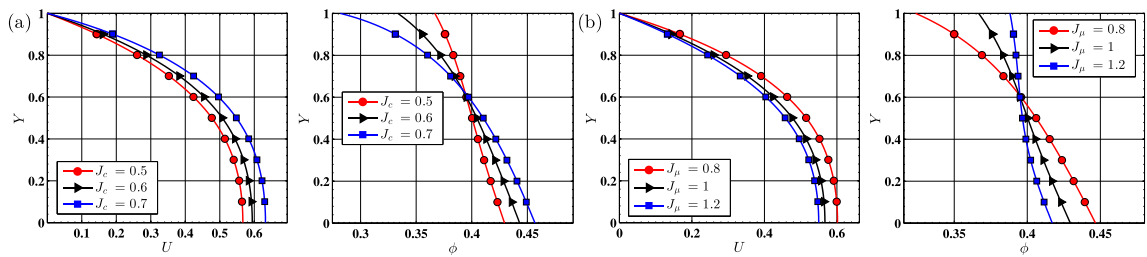

Figure 5. Parametric studies on the RBCs transport fluxes. (a) Effect of $J_c$ on the velocity profile (left) and the volume fraction profile (right), with $J_\mu = 1$. (b) Effect of $J_\mu$ on the velocity profile (left) and the volume fraction profile (right), with $J_c = 0.5$. For the above numerical studies, $\bar{k} = 10$, $B_{31} = 1$, $B_{32} = 5$, $\bar{\phi} = 0.4$, $\Delta P = 10$.

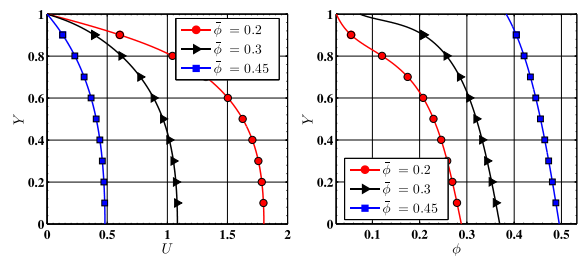

Figure 6. Effect of the hematocrit (averaged volume fraction) on the velocity profile (left) and the volume fraction profile (right), with $\bar{k} = 10$, $B_{31} = 1$, $B_{32} = 5$, $J_c = 0.65$, $J_\mu = 1$, $\Delta P = 10$.

## 6. Conclusions

As mentioned earlier, the multi-phase approach to study blood, due to the high computational cost has limited the use for many realistic engineering scale problems [25]. As alternative methods, multi-component models based on continuum mechanics, while ignoring the details of single cells, can overcome the limitations of high computational cost and still provide some useful information, for example, the volume fraction (hematocrit) and the velocity fields of RBCs and plasma [2] and the non-uniform distribution of platelet [26] can be obtained via the model developed in our current study.

In this paper, we introduce a non-linear suspension model for blood, where the viscosity depends on the local volume fraction and the shear rate. The migration of the RBCs is modeled using a concentration flux equation. The effects of gravity, viscosity and the shear rate are incorporated as RBCs fluxes in the concentration equation. The equations are non-dimensionalized and studied in two representative problems, namely a simple shear flow and a pressure driven flow between two plates. The numerical results indicate that for simple shear flow, in which the effect of gravity is included, the migration (sedimentation) of RBCs towards the lower plate is enhanced by the shear thinning parameter. For pressure driven flow, the RBCs tend to concentrate near the centerline, due to the flux term, $N_c$, where the shear rate is lowest. Both problems indicate that, in general, $N_c$ is responsible for the non-uniform distribution of the RBCs while the term $N_\mu$ plays the opposite role. The numerical results presented here are consistent with those predicted by our two-fluid method (mixture theory), which indicates that this simplified model retains key features of the two-fluid model, yet requires lesser computational cost.

## Acknowledgment

This research was supported by NIH grant 1 R01 HL089456.

## References

[1]     A.M. Robertson, A. Sequeira, M. V Kameneva, Hemorheology, in: Hemodynamical Flows, Springer, 2008: pp. 63–120.

[2]     W.-T. Wu, F. Yang, J.F. Antaki, N. Aubry, M. Massoudi, Study of blood flow in several benchmark micro-channels using a two-fluid approach., Int. J. Eng. Sci. 95 (2015) 49–59.

[3]     W.-T. Wu, M.A. Jamiolkowski, W.R. Wagner, N. Aubry, M. Massoudi, J.F. Antaki, Multi-Constituent Simulation of Thrombus Deposition, Sci. Rep. 7 (2017) 42720. doi:10.1038/srep42720.

[4]     Y.C. Fung, Biomechanics: Mechanical Properties of Living Tissues, 2nd ed., Springer, New York, 1993.

[5]     P. Bagchi, Mesoscale simulation of blood flow in small vessels, Biophys. J. 92 (2007)

1858–1877.

[6]     A.S. Popel, P.C. Johnson, Microcirculation and hemorheology, Annu. Rev. Fluid Mech. 37 (2005) 43.

[7]     H. Bäumler, B. Neu, E. Donath, H. Kiesewetter, Basic phenomena of red blood cell rouleaux formation, Biorheology. 36 (1999) 439–442.

[8]     S. Chien, Shear dependence of effective cell volume as a determinant of blood viscosity, Science (80-. ). 168 (1970) 977–979.

[9]     M.D. Rourke, A.C. Ernstene, A method for correcting the erythrocyte sedimentation rate for variations in the cell volume percentage of blood, J. Clin. Invest. 8 (1930) 545.

[10]    K.K. Yeleswarapu, M. V Kameneva, K.R. Rajagopal, J.F. Antaki, The flow of blood in tubes: theory and experiment, Mech. Res. Commun. 25 (1998) 257–262.

[11]    S. Middleman, Transport phenomena in the cardiovascular system, John Wiley & Sons, 1972.

[12]    S. Chien, S. Usami, R.J. Dellenback, C.A. Bryant, Comparative hemorheology--hematological implications of species differences in blood viscosity., Biorheology. 8 (1971) 35–57.

[13]    S. Chien, S. Usami, H.M. Taylor, J.L. Lundberg, M.I. Gregersen, Effects of hematocrit and plasma proteins on human blood rheology at low shear rates., J. Appl. Physiol. 21 (1966) 81–87.

[14]    D.E. Brooks, J.W. Goodwin, G. V Seaman, Interactions among erythrocytes under shear., J. Appl. Physiol. 28 (1970) 172–7.

[15]    R. Fahraeus, The suspension stability of the blood, Physiol. Rev. 9 (1929) 241–274.

[16]    R. Fahraeus, T. Lindvist, The viscosity of the blood in narrow capillary tubes, \Ajp. 96 (1931) 562–568.

[17]    C.K. Kang, A.C. Eringen, The effect of microstructure on the rheological properties of blood, Bull. Math. Biol. 38 (1976) 135–159. doi:10.1007/BF02471753.

[18]    G. Segré, a. Silberberg, Behaviour of macroscopic rigid spheres in Poiseuille flow Part 1. Determination of local concentration by statistical analysis of particle passages through crossed light beams, J. Fluid Mech. 14 (1962) 115. doi:10.1017/S002211206200110X.

[19]    G. Segré, a. Silberberg, Behaviour of macroscopic rigid spheres in Poiseuille flow Part 2. Experimental results and interpretation, J. Fluid Mech. 14 (1962) 136. doi:10.1017/S0022112062001111.

[20]    M.M. Dupin, I. Halliday, C.M. Care, L. Alboul, L.L. Munn, Modeling the flow of dense suspensions of deformable particles in three dimensions, Phys. Rev. E. 75 (2007) 66707.

[21]    J. Zhang, P.C. Johnson, A.S. Popel, Red blood cell aggregation and dissociation in shear flows simulated by lattice Boltzmann method, J. Biomech. 41 (2008) 47–55.

[22]    X. Li, A.S. Popel, G.E. Karniadakis, Blood–plasma separation in Y-shaped bifurcating microfluidic channels: a dissipative particle dynamics simulation study, Phys. Biol. 9 (2012) 26010.

[23]    D.A. Fedosov, B. Caswell, G.E. Karniadakis, A multiscale red blood cell model with accurate mechanics, rheology, and dynamics, Biophys. J. 98 (2010) 2215–2225.

[24]    K. Tsubota, S. Wada, T. Yamaguchi, Particle method for computer simulation of red blood cell motion in blood flow, Comput. Methods Programs Biomed. 83 (2006) 139–146.

[25]    M.A. van der Hoef, M. Ye, M. van Sint Annaland, A.T. Andrews, S. Sundaresan, J.A.M.

Kuipers, Computational Fluid Dynamics, Elsevier, 2006. doi:10.1016/S0065-2377(06)31002-2.

[26]   W.-T. Wu, N. Aubry, M. Massoudi, J.F. Antaki, Transport of platelets induced by red blood cells based on mixture theory, Int. J. Eng. Sci. 118 (2017) 16–27. doi:10.1016/j.ijengsci.2017.05.002.

[27]   K.R. Rajagopal, L. Tao, Mechanics of mixtures, Series on Advances in Mathematics for Applied Sciences, vol. 35, World Scientific, Singapore, 1995.

[28]   M. Ishii, Thermo-fluid dynamic theory of two-phase flow, Paris, Eyrolles, Ed. (Collection La Dir. Des Etudes Rech. d'Electricite Fr. No. 22), 1975. 275 P. 75 (1975).

[29]   W.-T. Wu, N. Aubry, M. Massoudi, J. Kim, J.F. Antaki, A numerical study of blood flow using mixture theory, Int. J. Eng. Sci. 76 (2014) 56–72.

[30]   C. Truesdell, Sulle basi della thermomeccanica, Rand Lincei. 8 (1957) 33-38 and 158-166.

[31]   M. Massoudi, A note on the meaning of mixture viscosity using the classical continuum theories of mixtures, Int. J. Eng. Sci. 46 (2008) 677–689.

[32]   R.J. Phillips, R.C. Armstrong, R.A. Brown, A.L. Graham, J.R. Abbott, A constitutive equation for concentrated suspensions that accounts for shear-induced particle migration, Phys. Fluids A Fluid Dyn. 4 (1992) 30–40.

[33]   W.-T. Wu, Z.-F. Zhou, N. Aubry, J.F. Antaki, M. Massoudi, Heat transfer and flow of a dense suspension between two cylinders, Int. J. Heat Mass Transf. 112 (2017) 597–606. doi:10.1016/j.ijheatmasstransfer.2017.05.017.

[34]   S.J. Hund, J.F. Antaki, An extended convection diffusion model for red blood cell-enhanced transport of thrombocytes and leukocytes., Phys. Med. Biol. 54 (2009) 6415–35.

[35]   K.R. Rajagopal, G. Saccomandi, L. Vergori, On the approximation of isochoric motions of fluids under different flow conditions, Proc. R. Soc. A. 471 (2015) 20150159.

[36]   W.-T. Wu, M. Massoudi, Heat Transfer and Dissipation Effects in the Flow of a Drilling Fluid, Fluids. 1 (2016) 4. http://www.mdpi.com/2311-5521/1/1/4/htm (accessed April 4, 2016).

[37]   K.R. Rajagopal, G. Saccomandi, L. Vergori, A systematic approximation for the equations governing convection–diffusion in a porous medium, Nonlinear Anal. Real World Appl. 11 (2010) 2366–2375.

[38]   C. Macosko, Rheology: Principles, Measurements and Applications, Wiley-VCH inc., New York, 1994.

[39]   K.R. Rajagopal, L. Tao, Mechanics of mixtures, World Scientific Publishers, Singapore, 1995. doi:10.1142/2197.

[40]   M. Massoudi, A Mixture Theory formulation for hydraulic or pneumatic transport of solid particles, Int. J. Eng. Sci. 48 (2010) 1440–1461.

[41]   A.C. Eringen, Continuum theory of dense rigid suspensions, Rheol. Acta. 30 (1991) 23–32. doi:10.1007/BF00366791.

[42]   A. Cemal Eringen, A continuum theory of dense suspensions, Zeitschrift Für Angew. Math. Und Phys. 56 (2005) 529–547. doi:10.1007/s00033-005-3119-2.

[43]   T. Ariman, M.A. Turk, N.D. Sylvester, Microcontinuum fluid mechanics—A review, Int. J. Eng. Sci. 11 (1973) 905–930. doi:10.1016/0020-7225(73)90038-4.

[44]   K.R. Rajagopal, G. Saccomandi, A novel approach to the description of constitutive relations, Front. Mater. 3 (2016) 36.

[45]  M. Anand, K.R. Rajagopal, A shear-thinning viscoelastic fluid model for describing the flow of blood, Int. J. Cardiovasc. Med. Sci. 4 (2004) 59–68.

[46]  G.B. Thurston, Viscoelasticity of human blood., Biophys. J. 12 (1972) 1205–17. doi:10.1016/S0006-3495(72)86156-3.

[47]  S. Chien, R.G. King, R. Skalak, S. Usami, A.L. Copley, Viscoelastic properties of human blood and red cell suspensions, Biorheology. 12 (1975) 341–346.

[48]  K.K. Yeleswarapu, Evaluation of continuum models for characterizing the constitutive behavior of blood., University of Pittsburgh, 1996.

[49]  G.B. Thurston, Frequency and shear rate dependence of viscoelasticity of human blood., Biorheology. 10 (1973) 375–381.

[50]  K.K. Yeleswarapu, J.F. Antaki, M. V Kameneva, K.R. Rajagopal, A generalized Oldroyd-B model as constitutive equation for blood, Ann. Biomed. Eng. 22 (1994) 16.

[51]  G.K. Batchelor, J.T. Green, The determination of the bulk stress in a suspension of spherical particles to order c 2, J. Fluid Mech. 56 (1972) 401–427.

[52]  A. Acrivos, R. Mauri, X. Fan, Shear-induced resuspension in a Couette device, Int. J. Multiph. Flow. 19 (1993) 797–802.

[53]  A. Mavromoustaki, A.L. Bertozzi, Hyperbolic systems of conservation laws in gravity-driven, particle-laden thin-film flows, J. Eng. Math. 88 (2014) 29–48.

[54]  M. Abedi, M.A. Jalali, M. Maleki, Interfacial instabilities in sediment suspension flows, J. Fluid Mech. 758 (2014) 312–326.

[55]  T. Hsu, P.A. Traykovski, G.C. Kineke, On modeling boundary layer and gravity-driven fluid mud transport, J. Geophys. Res. Ocean. 112 (2007).

[56]  L. Miao, M. Massoudi, Heat transfer analysis and flow of a slag-type fluid: Effects of variable thermal conductivity and viscosity, Int. J. Non. Linear. Mech. 76 (2015) 8–19.

[57]  K.R. Rajagopal, G. Saccomandi, L. Vergori, Flow of fluids with pressure-and shear-dependent viscosity down an inclined plane, J. Fluid Mech. 706 (2012) 173–189.

[58]  W.T. Wu, N. Aubry, M. Massoudi, Flow of granular materials modeled as a non-linear fluid, Mech. Res. Commun. 52 (2013) 62–68. doi:10.1016/j.mechrescom.2013.06.008.

[59]  J. Cao, G. Ahmadi, M. Massoudi, Gravity granular flows of slightly frictional particles down an inclined bumpy chute, J. Fluid Mech. 316 (1996) 197–221.

[60]  M.U. Guide, The mathworks, Inc., Natick, MA. 5 (1998) 333.

[61]  C. Bridges, K.R. Rajagopal, Pulsatile Flow of a Chemically-Reacting Nonlinear Fluid, Comput. Math. with Appl. 52 (2006) 1131–1144.

[62]  J.D. Humphrey, K.R. Rajagopal, A constrained mixture model for growth and remodeling of soft tissues, Math. Model. Methods Appl. Sci. 12 (2002) 407–430.